\documentclass[aps,amsmath,prl,twocolumn,superscriptaddress,showpacs]{revtex4}

\usepackage{graphicx}

\newcommand{\be}{\begin{equation}}
\newcommand{\ee}{\end{equation}}

\begin{document}


\title{Scalings of domain wall energies in two dimensional 
Ising spin glasses}

\author{C. Amoruso}
\affiliation{Dipartimento di Fisica, SMC and UdR1 of INFM, INFN, 
Universit\`a di  Roma {\em La Sapienza}, P.le Aldo Moro 2, 00185 Roma, Italy.}

\author{E. Marinari}
\affiliation{Dipartimento di Fisica, SMC and UdR1 of INFM, INFN, 
Universit\`a di  Roma {\em La Sapienza}, P.le Aldo Moro 2, 00185 Roma, Italy.}

\author{O. C. Martin}
\affiliation{Laboratoire de Physique Th\'eorique et Mod\`eles Statistiques,
b\^at. 100, Universit\'e Paris-Sud, F--91405 Orsay, France.}

\author{A. Pagnani}
\affiliation{Laboratoire de Physique Th\'eorique et Mod\`eles Statistiques,
b\^at. 100, Universit\'e Paris-Sud, F--91405 Orsay, France.}

\date{\today}

\begin{abstract}
We study domain wall energies of two dimensional spin glasses. The
scaling of these energies depends on the model's distribution of
quenched random couplings, falling into {\em three} different classes.
The first class is associated with the exponent $\theta \approx
-0.28$, the other two classes have $\theta = 0$, as can be justified
theoretically.  In contrast to previous claims we find that $\theta=0$
does not indicate $d=d_l^c$ but rather $d \le d_l^c$, where $d_l^c$ is
the lower critical dimension.
\end{abstract}

\maketitle

Spin glasses~\cite{MezardParisi87b} exhibit many subtle phenomena such
as diverging non-linear susceptibilities, aging and memory, making it
a real challenge to understand these materials. In spite of much work,
there is still no consensus even on the nature of the frozen order in
equilibrium.  More surprising still, the case of two dimensions also
is not completely understood. In particular, the scaling
of the stiffness, a cornerstone of spin glass theory, is different when the
spin-spin couplings are of the form $J_{ij}=\pm 1$ compared to 
when they have a Gaussian distribution~\cite{BrayMoore86}.
This has been confirmed since using more powerful
numerical techniques~\cite{KawashimaRieger97,HartmannYoung01},
and in fact 
it was interpreted in~\cite{HartmannYoung01} as a lack of universality,
but this is unexpected and unexplained. Here we solve this puzzle:
we find that different types of quenched disorder lead to three
distinct behaviors. In particular, we motivate why the class of models
that includes the case $J_{ij}=\pm 1$ gives for the
stiffness exponent $\theta = 0$, and we explain
what $\theta$ tells us about the lower critical dimension.

\paragraph*{The model, its properties and our methods ---} 
The model consists of $N=L^2$ Ising spins $S_i = \pm 1$ on a simple square
lattice with periodic boundary conditions. The Hamiltonian is 
\begin{equation}
\label{eq:hamilton}
H \equiv -\sum_{\langle i j \rangle} S_i J_{ij} S_j\,,
\end{equation}
where the sum runs over all pairs of nearest neighbors $\langle i j
\rangle$ and the $J_{ij}$ are the quenched random spin-spin couplings.
We shall consider different distributions of these couplings, all of
which are symmetrical about $J=0$. We begin with continuous
distributions; most common is the one where the $J_{ij}$ are Gaussian
random variables with zero mean and unit variance.  After that we
investigate discrete distributions; the most common distribution of
this type has $J_{ij} = \pm 1$ with equal probability.

An important feature of spin glass ordering is the spin glass
stiffness; the corresponding exponent $\theta$ describes how
excitation free energies scale with the associated length scale. The
standard way to measure this exponent is via the change in the
system's free energy when going from periodic to anti-periodic
boundary conditions. At $T=0$ this reduces to measuring the difference
\begin{equation}
\label{eq:deltaE}
\delta E = E_0^{(P)} - E_0^{(AP)}\,,
\end{equation} 
where $E_0^{(P)}$ and $E_0^{(AP)}$ are the ground state energies for the
system with respectively periodic and anti-periodic boundary
conditions say in the $x$ direction.  We are interested in the
probability distribution of $\delta E$ when considering an ensemble of
$J_{ij}$ and in the scaling law of its standard deviation $\Delta E$:
\begin{equation}
\Delta E\underset{L \to \infty}{\sim} ~ L^{\theta}\,.
\end{equation}  
Measurements of $\theta$ in two dimensional spin glasses (see for
instance~\cite{BrayMoore86}) give $\theta \approx -0.28$. However,
for the $J_{ij}=\pm 1$ distribution, Hartmann and 
Young~\cite{HartmannYoung01} recently showed that
$\Delta E$ remains of $O(1)$ for increasing $L$,
implying that in this case $\theta \approx 0$.
In dimension $d$ above the lower critical dimension $d_l^c$ we have
$\theta > 0$ and spin glass ordering is stable against thermal
fluctuations.  On the contrary, when $\theta < 0$, thermal
fluctuations prevent spin glass ordering.  Because of this,
the authors of~\cite{HartmannYoung01} conjectured that $d_l^c = 2$
for the $J_{ij}=\pm 1$ model.  We shall see that $d_l^c$ should be
identified with the highest value of $d$ where $\theta \le 0$, and so
in fact $d_l^c \approx 2.5 $ as believed before the study 
in~\cite{HartmannYoung01}.
 
In this work we address these questions by first determining
numerically the properties of $P(\delta E)$ and then by using the real
space renormalization group picture. For the first part, we compute
the ground states of our systems using 
a heuristic algorithm~\cite{HoudayerMartin01}.  
In practice, when the lattice is not too large ($L \le 80$), the
algorithm returns the ground state with a high level of confidence for
all of the distributions we shall consider in this work. The problem
is to reduce enough the statistical errors; in practice we used a few
tens of thousands of samples at a few values of $L$ for each case.

\paragraph*{Class 1: ``continuous'' distributions ---} 
We first focus on distributions $P(J)$ that include a continuous part
(we shall see later that this class includes certain discrete
distributions also). When $L$ is sufficiently large, $\delta E$ can
then take on arbitrary values.  The value of $\theta$ for continuous
distributions is well known only for Gaussian $J_{ij}$; in fact we are
aware of no tests of universality in $d=2$, though the
standard lore is that both $\theta$ and the shape of $P(\delta E)$ are
universal~\cite{BrayMoore86}.

In a first series of runs we obtained $P(\delta E)$ and
$\Delta E$ for the model with Gaussian couplings. Then we moved on to
a continuous yet singular probability density $P(J_{ij})$:
$  P(J_{ij}=J)$ $=$ $f ~ P_1(J) + (1-f) ~ P_2(J)$,
where
$P_1(J)$   $\equiv  [ e^{\frac{{(J-1)}^2}{2}}
                         +e^{\frac{{(J+1)}^2}{2}} ] /
\sqrt{8\pi}$, 
$P_2(J)$  $\equiv [ \delta(J - 1) +\delta(J + 1) ] / 2$,
and $f$ is a measure of the height of the distribution 
at $J \approx 0$. We refer to this $P(J_{ij})$ as the
{\em broadened bimodal (BB) distribution} since it reduces
to the $J_{ij}= \pm 1$ distribution when $f=0$. 

\begin{figure}[htb]
\includegraphics[angle=0,width=\columnwidth]{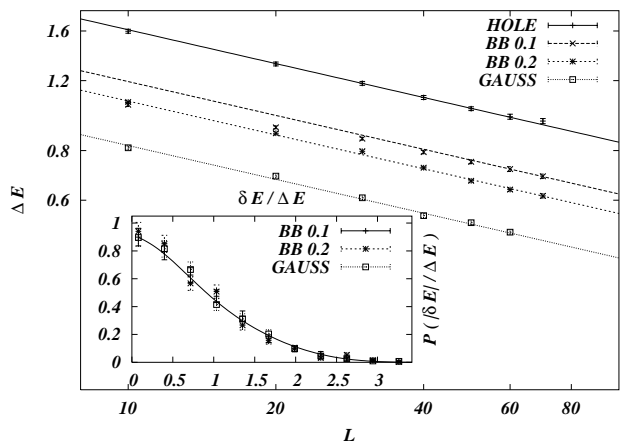} 
\caption{$\Delta E$ as a function of the system size for four
different $J_{ij}$ distributions. Straight lines are best
one-parameter fits of the form $const \times L^{-0.282}$.  Inset:
the probability distribution $P(|\delta E| / \Delta E)$ at $L=40$ for
three of these distributions.}
\label{fig:delta_E_continuous}
\end{figure}

In Fig. \ref{fig:delta_E_continuous} we show $\Delta E$ as a
function of $L$ when $P(J_{ij})$ is: (1) a Gaussian of zero mean and
unit variance ($GAUSS$ data); (2) the BB distribution,
with $f = 0.1$ ($BB$ $0.1$ data); (3) as in (2) but with $f = 0.2$
($BB$ $0.2$ data); (4) Gaussian but with the part in the interval
$[-0.5,0.5]$ forced to be $0$ ($HOLE$ data). Note that this last distribution
has a large gap around $J_{ij}=0$.
In the Gaussian case the power law scaling of $\Delta E$ can be
determined with good accuracy already from quite small lattices; fits
to these data lead to $\theta=-0.282 \pm 0.004$, in agreement
with previous work. The distributions (2), (3) and
(4) give rise to a similar scaling albeit only at larger $L$
values. We have also considered other distributions such as
$P(J_{ij})$ uniform in $[-1.5,-0.5] \cup [0.5,1.5]$ (notice that this
distribution also has a gap around $J=0$), obtaining similar results. It
thus seems very reasonable to expect that {\em all} distributions with
a continuous part will lead to the same 
exponent, $\theta \approx -0.28$.

A second universality issue concerns the {\em shape} of
$P(\delta E)$.  In the inset of Fig. \ref{fig:delta_E_continuous}
we show the probability density $P(|\delta E| / \Delta E)$ when $L=60$
for the $BB$ $0.1$, $BB$ $0.2$, and $GAUSS$ data: the different data
sets basically coincide within statistical errors,
strengthening the claim that in this class the distribution of domain wall
energies is universal (the curve displayed is just to guide the eye).

\begin{figure}[htb]
\includegraphics[angle=0,width=\columnwidth]{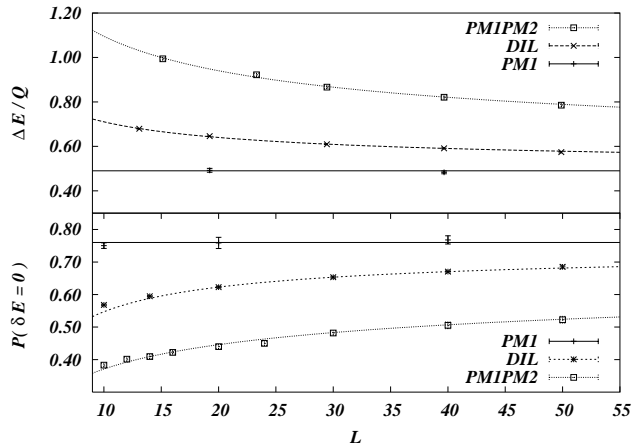} 
\caption{$\Delta E / Q$ (top) and $P(\delta E = 0)$ (bottom) as a
function of the system size for three discrete $J_{ij}$ distributions:
$\pm 1$ (PM1), diluted $\pm 1$ (DIL), and $\pm 1$, $\pm2$ (PM1PM2).}
\label{fig:delta_E_discrete}
\end{figure}

\paragraph*{Class 2: quantized energies ---} 
At variance with the former distributions, the $J_{ij}=\pm 1$ model
leads to $\theta \sim 0$~\cite{HartmannYoung01}.  We show in Fig.
\ref{fig:delta_E_discrete} that in this model $\Delta E$ saturates
quickly as $L$ grows.  Is the $J_{ij} = \pm 1$ model a special case, a
class on its own?  The crucial point is that the possible $\delta E$
values are {\em quantized}: $\delta E$ is always a multiple of a {\em
quantum} $Q$, here $Q=4$. This led us to consider distributions other
than the $\pm 1$ one with this same quantization property.  We begin
by ``diluting'' the $J_{ij} = \pm 1$ model, setting $J_{ij}=0$ with
probability $0.2$. The main effect of this is to reduce the quantum
from $4$ to $2$; indeed, the local fields now can take the value
0,1,2,3,4 instead of 0,2,4. In Fig. \ref{fig:delta_E_discrete} we
see that for this model ($DIL$ data) $\Delta E$ seems to saturate, so
again $\theta = 0$.  However the convergence is slow. In any
renormalization group picture this convergence is governed by a
``correction to scaling'' exponent $\omega$. We assume $\theta=0$ and
that the asymptotic value of $\Delta E$ is a non-zero constant given
by the $J=\pm1$ data; then we fit the diluted (DIL) model to the form:
\begin{equation}
\Delta E (L) \approx \Delta E(L=\infty) + A L^{-\omega} \ .
\label{eq:convergence}
\end{equation}
with $A$ and $\omega$ adjustable parameters. We have also considered
distributions where $J_{ij} = \pm J_1$ or $\pm J_2$ with equal
probability (we have studied the cases $J_2/J_1 = 1.5$, $2$ and
$3$.). Again we find the convergence to be slow but fits
as in (\ref{eq:convergence}) work well; furthermore,
all the estimates of $\omega$ are similar, being in the
$[0.4:0.6]$ interval. All these facts justify
the claim that $\theta = 0$ whenever $\delta E$ is quantized.

Just as in the continuous case, to analyze the {\em shape} of the 
distribution of $\delta E$ we must choose a scale; the correct choice
is to compare the histograms after measuring all energies in units
of the basic quantum $Q$.  To test whether the histograms for the
different $J_{ij}$ distributions become identical in the large $L$
limit we plot in Fig. \ref{fig:delta_E_discrete} (lower panel) the
probability $P(\delta E = 0)$ to find a zero energy domain wall. The
data suggest that the histograms become identical in the
large $L$ limit, i.e., they support universality. (Following
Eq.~(\ref{eq:convergence}), we fix 
the asymptotic value of $P(\delta E = 0)$ to 
be that given by the $J=\pm 1$ model, and then we determine
$\omega$; in the plot we show these fits; they are
all good and the values of $\omega$ are close to $0.5$.) We 
have checked in detail that this claim applies to
the quantized distributions mentioned before {\em and} to the $DIL$
model with $10\%$ dilution.

\paragraph*{Class 3: quantized energies revisited ---}
So far we have only considered situations with even values of $L$.
If $L$ is odd (and the $J_{ij}=\pm1$), the possible values 
of $\delta E / J$ are $\pm 2, \pm 6, \pm 10$, \ldots The quantum $Q$ is
still the separation between the energy values, but the positions of
the histogram entries are different (in particular, $\delta E = 0$ is
not allowed).  A somewhat trivial consequence of this is that
necessarily $\theta \ge 0$ as $\Delta E/Q$ is greater or equal to
$1/2$ for all $L$. Consider now the question of the universality of
the histograms.  We have checked within our error bars that the large
$L$ limit of $P(\delta E / Q)$ for the $J_{ij}=\pm1$ model is the same
as that obtained using the $J_2/J_1=3$ model (still with $L$ odd of
course). This kind of quantization thus gives rise to a third class,
again with $\theta = 0$.

Could there be further classes with quantized energies?
Since we have imposed reflection symmetry of the distribution of the
$J_{ij}$ the only possible histograms are the two we discussed:
$\delta E$ is a multiple of the quantum $Q$ or of the form $(n+1/2)Q$,
where $n$ is integer. If the universality class depends only on the
possible histogram types, then no other classes arise.

\paragraph*{Discrete does not mean quantized ---} 
Let us also consider the case where the couplings are {\em discrete} 
but where there is no quantization. We consider the
distribution
$P(J)= \frac{1}{4}[ \delta(J \pm J_1 ) +\delta(J \pm J_2)]$
(IRR for ``irrational'' hereafter), where $J_1 = 1$ and $J_2 =
\frac{1+\sqrt{5}}{2} \approx 1.618$ is the golden mean. 
Clearly we have $\delta E = 2 (n J_1+ m J_2)$
where $n$ and $m$ are integers.  Since $J_2/J_1$ is irrational, the
set of possible $\delta E$ values becomes dense when $L \to \infty$
and so it is natural to conjecture that
this $P(J_{ij})$ leads to domain wall energies in class 1.
Our findings are that $\Delta E$ decreases with $L$ and shows
no sign of saturation, and a power law fit gives $\theta = -0.29 \pm
0.01$, the value associated with class 1. Our
conjecture is thus substantiated by these findings.

\begin{figure}[htb] 
\includegraphics[angle=0,width=\columnwidth]{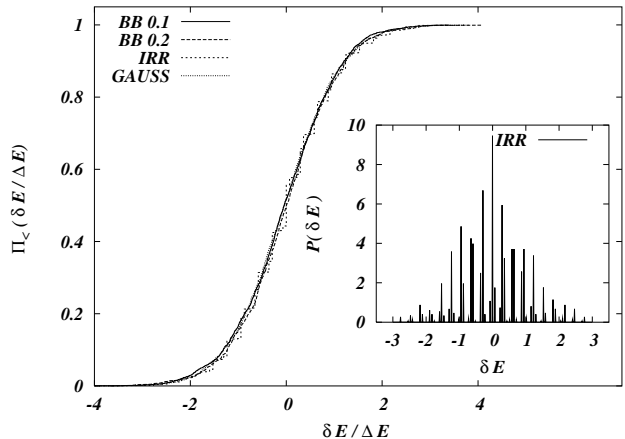}
\caption{The integrated probability distribution for $\delta E/\Delta
E$ in the Gaussian case, in the two BB cases, and in the
irrational $J_2/J_1$ case ($L=60$). Inset: binned probability
distribution of $\delta E / \Delta E$ for the irrational case
(IRR), $L=60$.}
\label{fig:irrational}
\end{figure}

The convergence of $P(\delta E / \Delta E)$ to its limit is more
problematic: for finite $L$, the distribution is the sum of a finite
number of delta functions: we can only hope to have a ``weak convergence''
to the $P$ obtained with the Gaussian couplings.  In these conditions
it is appropriate to consider the integrated probability distribution
$\Pi_< (X) \equiv \int_{-\infty}^{X} P\Big({\frac{\delta E}{\Delta
E}}\Big) d \left(\frac{\delta E}{\Delta E}\right)$.
In Fig. \ref{fig:irrational} we plot
$\Pi_<$ for the irrational and for some continuous cases.  The
plots are very similar, supporting the claim that the discrete
distribution IRR leads to domain wall energies in class 1.
We have also included $P(\delta E / \Delta E)$ as an inset 
into Fig. \ref{fig:irrational}: we have used a small bin size that
allows one to observe the complex structure.

\paragraph*{The case of hierarchical lattices} --- The effect of having 
quantized $\delta E$ can also be studied on hierarchical
lattices. One advantage is that one can study very large sizes, a
second is that one can access a {\em continuous} range of dimensions.
We have focused on Migdal-Kadanoff lattices \cite{SouthernYoung77};
these are obtained by recursively ``expanding'' graphs. Starting with
one edge connecting two sites, one replaces it by $b$ paths in
parallel, each composed of $s$ edges in series, leading to $b\cdot s$
new edges. This procedure is repeated hierarchically;
after $G$ ``generations'' the distance
between the outer-most spins is $L = s^G$, while the number of edges
of the lattice is $(b\cdot s)^G$. The dimension of these lattices is
$d = 1 + \ln(b) / \ln(s)$.  One puts an Ising spin on each site and a
coupling $J_{ij}$ on each edge. Periodic boundary conditions simply
imply that the two end spins must have the same value, and from this
we define $\delta E$. The probability distribution of $\delta E$ can
be followed from $G$ to $G+1$.
The recursion equations for $P(\delta E)$ make sense for any
$s$~\cite{CuradoMeunier88}: $s$ can be an integer but it can also 
be any positive real
value! One may then compute $\theta$ for an interval of dimensions,
using either continuous $J_{ij}$
(for instance to check universality~\cite{NogueiraCoutinho98,Boettcher03d})
or quantized $J_{ij}$ couplings (our focus here).

\begin{figure}[htb] 
\includegraphics[angle=0,width=\columnwidth]{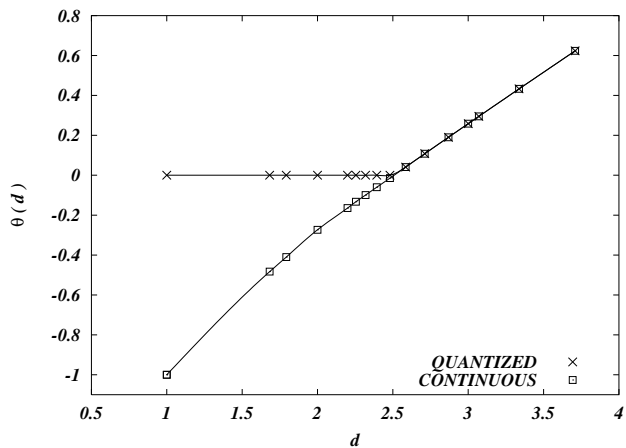}
\caption{$\theta$ as a function of $d$ for a
one-parameter family of Migdal-Kadanoff lattices; we display two sets
of data points, one for continuous $J_{ij}$ distributions, the other for
quantized distributions.}
\label{fig:thetaMK}
\end{figure}

In Fig.~\ref{fig:thetaMK} we show $\theta$ as a function of
dimension $d$ ($s$ is variable, $b$ is fixed and set to $3$).  We show
the values for continuous distributions and for when the quantization
is of the form of class $3$. As expected, if in one class $\theta>0$,
all classes lead to the same value of $\theta$, i.e., quantization is
irrelevant when the energy scale diverges.  However, as soon as
$\theta < 0$ in the continuous case, quantization gives rise to a
histogram fixed point distribution in which the $| \delta E |$ are
concentrated on the few lowest values and $\theta = 0$.

Similar results are obtained for class 2 quantization but there
is an interesting difference. Indeed, since
$\delta E$ can be zero in this class,
one sees two further fixed points. An obvious one
is associated with having $P(\delta E = 0) = 1$, 
i.e,. all domain wall energies vanish. It is easy to see that this fixed
point is stable and has $\theta = -\infty$; there
is no spin glass stiffness, and the system is paramagnetic even
at zero temperature. The other fixed point is unstable and 
has $\theta = 0$.
What is the interpretation of these two extra fixed points? To allow
$\delta E$ to be zero, one can think of the diluted model where some
of the bonds have $J_{ij}=0$. Clearly when the dilution is strong
enough, the non-zero bonds will no longer percolate and we are in a
strongly paramagnetic phase; the renormalization group (RG) 
flow in this phase takes one to
the $P(\delta E = 0) = 1$ fixed point.  On the contrary, at low
dilution, we are in a spin glass phase and the RG flows are towards
the other stable fixed point.  On the {\em boundary} of these two phases,
the RG flows takes one to another fixed point which is unstable: it is
associated with the paramagnetic to spin glass transition as dilution
is decreased. Such considerations have previously been developed 
for $d=3$ Migdal-Kadanoff lattices~\cite{BrayFeng87}.

Finally, we see that it is appropriate to define the lower critical
dimension $d_l^c$ from the end point of the $\theta=0$ curve; $\theta=0$ on
its own does {\em not} signal $d = d_l^c$.

\paragraph*{Discussion ---} 
Our numerical evidence of universality (both for $\theta$ and
$P(\delta E / \Delta E)$) is very strong for continuous and related
distributions (see figures
\ref{fig:delta_E_continuous} and \ref{fig:irrational}). But we also find
universality classes when $\delta E$ is quantized. This classification
is substantiated by the behavior of $\theta$ and of the fixed point
distributions of domain wall energies in Migdal-Kadanoff lattices.  It
is appropriate however to be cautious and to remark that the
correction to scaling exponent $\omega$ we measure (see 
equation~\ref{eq:convergence}) is small, $\omega \approx 0.5$. Because of that
we are not able to completely exclude the Bray and Moore expectation
that $\Delta E(L = \infty) = 0$ \cite{BrayMoore87}. Our most extensive
data are for the model DIL with $f=0.2$. Here our fits give $\Delta
E(L=\infty) / Q = 0.49(1)$ while if we force $\Delta E(L=\infty) = 0$,
the $\chi^2$ of the fit increases by $2.3$; thus $\Delta E(L=\infty) =
0$ is not excluded by our data though it appears as much less likely.

What is the source of the universality we observe? In the
Migdal-Kadanoff lattices, the renormalization group transformation is
clear and so the different classes are very natural. For the
Euclidean lattices the existence of a renormalization group
transformation for $\delta E$ has not been established, but since our
data point to universality, it should be possible to define such a
transformation.  Note that its fixed point (and thus $P(\delta E /
\Delta E)$) will depend on the aspect ratio and on the fact that we
use periodic boundary conditions. Our $P(\delta E)$ are thus {\em a
priori} not comparable to those of~\cite{HartmannYoung01} where one of
the directions had free boundary conditions.

We acknowledge important conversations with Giorgio Parisi that led us
to this study, and we thank J.-P. Bouchaud for his comments. AP
acknowledges the financial support provided through the European
Community's Human Potential Programme under contract
HPRN-CT-2002-00307, DYGLAGEMEM.

\bibliographystyle{apsrev}

\bibliography{references}

\end{document}